\documentclass[journal]{IEEEtran}

\ifCLASSINFOpdf
\else
\fi

\usepackage{graphics, epsf,psfrag, subfigure,epsfig, times,amsmath,amssymb,color}
\usepackage{url}
\usepackage{verbatim} 
\usepackage{multicol}
\usepackage{array}
\usepackage{margins}
\usepackage{psfrag}
\usepackage{cite}

\usepackage{amsthm}
\usepackage{amsmath, amssymb}
\usepackage{amsfonts}
\newcommand{\bfb}{{\mathbf b}}

\newcounter{actr}
{\begin{list}{(\alph{actr})}{\usecounter{actr}}}{\end{list}}

\newcounter{ictr}
{\begin{list}{(\roman{ictr})}{\usecounter{ictr}}}{\end{list}}

\newtheorem{remark}{Remark}
\newtheorem{thm}{Theorem}
\newtheorem{lemma}{Lemma}

\newtheorem{prop}{Proposition}

\newenvironment{new-proof}
{{\em Proof: }}
{ \noindent\qed }

\renewcommand{\qed}{\rule[0.1ex]{1.4ex}{1.6ex}}

\newcommand{\mrm}{\mathrm}

\newcommand{\cA}{{\mathcal{A}}}

\newcommand{\bb}{{\mathbf{b}}}

\newcommand{\cB}{{\mathcal{B}}}

\newcommand{\cC}{{\mathcal{C}}}

\newcommand{\cD}{{\mathcal{D}}}

\newcommand{\cE}{{\mathcal{E}}}

\newcommand{\bg}{{\mathbf{g}}}

\newcommand{\bG}{{\mathbf{G}}}

\newcommand{\cS}{{\mathcal{S}}}

\newcommand{\cT}{{\mathcal{T}}}
\newcommand{\bu}{{\mathbf{u}}}

\newcommand{\bw}{{\mathbf{w}}}

\newcommand{\bx}{{\mathbf{x}}}

\newcommand{\by}{{\mathbf{y}}}

\newcommand{\bz}{{\mathbf{z}}}

\newcommand{\g}{\gamma}

\newcommand{\del}{\delta}

\newcommand{\eps}{\varepsilon}

\DeclareMathAlphabet{\mathbsf}{OT1}{cmss}{bx}{n}
\DeclareMathAlphabet{\mathssf}{OT1}{cmss}{m}{sl}

\DeclareSymbolFont{bsfletters}{OT1}{cmss}{bx}{n}
\DeclareSymbolFont{ssfletters}{OT1}{cmss}{m}{n}
\DeclareMathSymbol{\bsfGamma}{0}{bsfletters}{'000}
\DeclareMathSymbol{\ssfGamma}{0}{ssfletters}{'000}
\DeclareMathSymbol{\bsfDelta}{0}{bsfletters}{'001}
\DeclareMathSymbol{\ssfDelta}{0}{ssfletters}{'001}
\DeclareMathSymbol{\bsfTheta}{0}{bsfletters}{'002}
\DeclareMathSymbol{\ssfTheta}{0}{ssfletters}{'002}
\DeclareMathSymbol{\bsfLambda}{0}{bsfletters}{'003}
\DeclareMathSymbol{\ssfLambda}{0}{ssfletters}{'003}
\DeclareMathSymbol{\bsfXi}{0}{bsfletters}{'004}
\DeclareMathSymbol{\ssfXi}{0}{ssfletters}{'004}
\DeclareMathSymbol{\bsfPi}{0}{bsfletters}{'005}
\DeclareMathSymbol{\ssfPi}{0}{ssfletters}{'005}
\DeclareMathSymbol{\bsfSigma}{0}{bsfletters}{'006}
\DeclareMathSymbol{\ssfSigma}{0}{ssfletters}{'006}
\DeclareMathSymbol{\bsfUpsilon}{0}{bsfletters}{'007}
\DeclareMathSymbol{\ssfUpsilon}{0}{ssfletters}{'007}
\DeclareMathSymbol{\bsfPhi}{0}{bsfletters}{'010}
\DeclareMathSymbol{\ssfPhi}{0}{ssfletters}{'010}
\DeclareMathSymbol{\bsfPsi}{0}{bsfletters}{'011}
\DeclareMathSymbol{\ssfPsi}{0}{ssfletters}{'011}
\DeclareMathSymbol{\bsfOmega}{0}{bsfletters}{'012}
\DeclareMathSymbol{\ssfOmega}{0}{ssfletters}{'012}

\newcommand{\rvb}{{\mathssf{b}}}    

\newcommand{\rvs}{{\mathssf{s}}}    

\newcommand{\rvu}{{\mathssf{u}}}    

\newcommand{\rvw}{{\mathssf{w}}}    

\newcommand{\rvx}{{{\mathssf{x}}}}    

\newcommand{\rvy}{{\mathssf{y}}}    

\newcommand{\rvz}{{\mathssf{z}}}    

\begin{document}

\title{On Modulo-Sum Computation over  an\\ Erasure Multiple Access Channel}

\author{Ashish Khisti,~\IEEEmembership{Member IEEE}, Brett Hern, and Krishna Narayanan,~\IEEEmembership{Senior Member IEEE}\thanks{ Ashish Khisti is with the University of Toronto, Toronto, ON, Canada email: akhisti@comm.utoronto.ca.  Brett Hern and Krishna Narayanan are with
Texas A\&M University, College Station Texas. Email: \{krn@tamu.edu, hernbrem@neo.tamu.edu\}. Ashish Khisti's work was supported by a  Discovery Research Grant from National Science Engineering Research Council (NSERC), Canada and Helwett-Packard Innovation Research Proposal  (HP-IRP) Award. Brett Hern and Krishna Narayanan were supported by the National Science Foundation under Grants
CCF 0729210 and 0830696. Part of this work will be presented at the 2012 International Symposium on Information Theory (Boston, MA).}}


\maketitle

\begin{abstract}
We study computation of a modulo-sum of two binary source sequences over a two-user erasure multiple access
channel. The channel is modeled as a binary-input, erasure multiple access channel, which can be in one of three states - either the channel output is a modulo-sum of the two input symbols, or the channel output equals the input symbol on the first link and an erasure on the second link, or vice versa. The associated state sequence is independent and identically distributed. We develop a new upper bound on the sum-rate by revealing only part of  the state sequence to the transmitters.  Our coding scheme is based on the compute and forward and the decode and forward techniques. When a (strictly) causal feedback of the channel state is available to the encoders, we show that the modulo-sum capacity is increased. Extensions to the case of lossy reconstruction of the modulo-sum and to channels involving additional states are also treated briefly.
\end{abstract}

\IEEEpeerreviewmaketitle


\vspace{-1em}
\section{Introduction}

In many emerging  applications in networked systems, it is sufficient for  intermediate nodes to compute a function of the  source messages. For example in a  two-way relay channel,  the two users need to mutually exchange messages using a central relay node. It is natural that the relay node only computes a modulo-sum of the  messages.  In other applications, the destination node may only be interested in some pre-determined function of the observations made by remote terminals. For example, in a temperature monitoring system, the fusion centre may only be interested in computing an average of the observations made by each of the sensor nodes.

Korner and Marton~\cite{KornerMarton:79} introduce a multi-terminal source coding problem where the destination terminal is required to compute a modulo-sum of two binary sources. Each source is revealed to one encoder and the source sequences need to be compressed such that the destination can recover the modulo-two sum of the two binary source sequences. 
The authors establish the optimality of a scheme that uses identical linear codebooks for compressing the two source sequences. There has been a significant interest in both source  and channel coding techniques for in-network function computation in recent times; see e.g.,~\cite{yamamoto:82, feng:04, doshi:07, pradhan, Wilson,Nazer2, NazerGastpar:11, sahebi, zamir, hern, agrawal,He:Structured,oechtering,zhang,philosof}.

We study the computation of a modulo-sum of two messages over a multiple access channel, introduced in ~\cite{Wilson,Nazer2}. These works consider
the Gaussian multiple access channel (MAC) and observe that for a wide range of signal-to-noise ratio (SNR), one can achieve higher rates using lattice codes instead of an i.i.d.\ random  code ensemble. 
Because of its  additive nature, the Gaussian MAC channel is well suited for computing the modulo sum of two messages using lattice codes.  
A simple  upper bound, obtained by revealing one of the messages  to the destination, suffices to establish the near-optimality of lattice-based schemes for a wide range of channel parameters. Similar schemes can also be developed for computation of a modulo-sum over the binary multiple-access channel. 

 In the present paper we study a  MAC channel model that does not appear naturally matched for computing the modulo-sum function. 
 Our model is an  erasure multiple access channel with binary inputs.
 With a certain probability, the destination observes a modulo-sum of the two transmitted bits whereas with a certain probability the destination observes only one of the two bits and an erasure symbol associated with the other transmitted bit. We establish upper and lower bounds on the modulo sum capacity of such a channel model. 
 The upper bound is tighter than the simple upper bound obtained by revealing one of the messages to the destination.
 The  lower bound is based on compute-and-forward and decode-and-forward schemes used in earlier works.
 It can be achieved by using identical linear codebooks at the two senders.  
 We also briefly consider the case when there is strictly causal feedback of the state sequence available from the destination (using e.g., ARQ) 
 and show that the capacity can be increased compared to the case without such feedback. 

Erasure channel models are suitable when one considers error-control coding in the upper layers of the protocol stack. A system could be designed such that
when both the transmitting nodes are active, the physical layer computes the modulo sum of the information bits and passes it to the upper layer.
Due to back-off mechanisms a transmitting node may not be active in each slot.  This leads to erasures on the respective links as considered in this paper.

\section{Problem Statement}
We study a multiple access channel with two transmitters and one receiver. The channel input symbols are denoted by $\rvx$ and $\rvy$ respectively and are binary valued. The channel output is denoted by $\rvz$ and is also binary valued. The channel transition probability is controlled by a state variable $\rvs \in \{0,1,2\}$. In particular we have:\begin{align}
\rvz =\begin{cases}
\rvx \oplus \rvy, & \rvs = 0,\\
\rvx, & \rvs = 1,\\
\rvy, & \rvs = 2.
\end{cases}\label{eq:chModel}
\end{align}
We assume that the receiver is revealed the pair $(\rvz, \rvs)$. 
We assume that $\Pr(\rvs=1)=\Pr(\rvs=2) = \eps$ and $\Pr(\rvs=0)=1-2\eps$ where $\eps$  satisfies $0 \le \eps \le 1/2$. 
The channel is memoryless i.e., $\Pr(\rvs^n = s^n) = \prod_{i=1}^n \Pr(\rvs_i = s_i)$. 

A code of length $n$ is defined as follows. 
Sender $i$ observes a message $\rvw_i$ uniformly and independently distributed over the set $[1,\ldots, 2^{nR}]$. 
For sake of convenience we will represent message $\rvw_i$ as a sequence $b_i^{nR}$ consisting of $nR$ independent and equiprobable bits.
We define ${\rvu = \rvw_1 \oplus \rvw_2}$ as the exclusive-or of $b_1^{nR} \oplus b_2^{nR}$. 

The messages are mapped into codewords $\rvx^n = f_n(\rvw_1)$ and $\rvy^n=g_n(\rvw_2)$ respectively and the decoder is required to produce ${\hat
{\rvu} = h_n(\rvz^n, \rvs^n)}$. An error is declared if $\{\rvu \neq \hat{\rvu}\}$.  

A rate $R$ is achievable if there is a sequence of encoders and decoders such that the error probability goes to zero as $n$ approaches infinity. The largest achievable rate is defined as the {\em modulo-sum capacity}.

\section{Main Results}
We state the main results in this section. 
\subsection{Lower Bound}
We propose the following lower bound on the modulo-sum capacity.
\begin{prop}
\label{prop:c_lb}
The modulo-sum capacity is lower bounded by the following expression:
\begin{equation}
C \ge R^- = \max\left\{1-2\eps, \frac{1}{2}\right\}.
\label{eq:c_lb}
\end{equation}
\end{prop}
The lower bound of ${R = 1-2\eps}$ is attained using a compute-and-forward technique~\cite{Nazer2} where identical linear codebooks are used by the two transmitters. The lower bound $R=1/2$ can be attained in several ways. Perhaps the simplest way is to transmit $\rvw_1$ and $\rvw_2$ to the destination using independent multiple-access channel codebooks\cite{coverThomas}. We call this scheme decode-and-forward. Interestingly if we use identical codebooks at the two transmitters~\cite{hern} for decode-and-forward, the rate $R=\min(1/2,2\eps)$ is achieved. As we will show,  
a variant of the compute-and-forward scheme also achieves $R=1/4$, when $\eps > 1/4$.

\subsection{Upper Bound}

We provide the following upper bound on the modulo-sum capacity.

\begin{figure}
\includegraphics[scale=0.35]{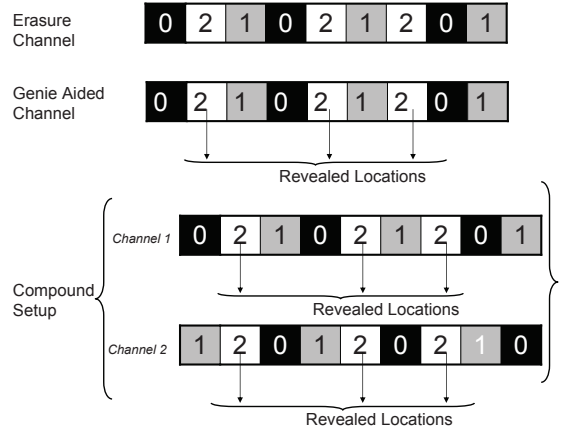}
\caption{Main Steps in the Upper Bound (for $\eps=1/3)$. The uppermost figure illustrates the erasure MAC model. Each square corresponds to one channel use. 
The black squares  correspond to $\rvs_i = 0,$ i.e., $\rvz_i = \rvx_i\oplus \rvy_i$, the shaded grey squares correspond to $\rvs_i = 1,$ i.e., $\rvz_i = \rvx_i$ and the white squares correspond to $\rvs_i=2$ i.e., $\rvz_i = \rvy_i$. Our upper bound reveals the location of $\rvs_i=2$ to both the transmitters  non-causally.  Since the transmitters are not aware of the location of the grey and black squares,
any code for this genie-aided channel must also be decodable when the black and grey squares are interchanged. This compound setup results in  a tighter upper bound than the usual cut-set bound.}
\label{fig:ub}
\end{figure} 

\begin{thm}
\label{thm:c_ub}
The modulo-sum capacity is upper bounded by the following expression:
\begin{equation}
C \le R^+ = \frac{(1-3\eps)^+ + (2-\eps)}{3}
\label{eq:c_ub}
\end{equation}
where $(\cdot)^+$ equals zero if the argument inside is negative. 
\end{thm}

The proposed upper bound is tighter than a genie-aided bound where  one of the messages, say $\rvw_1$, is revealed to the decoder. 
We provide the key-steps in the upper bound derivation below.

\subsubsection{Revealing Side Information to the Transmitters} 
Our key step is to  reveal part of the state sequence to the encoders. In particular define the sets $\cA = \{i: \rvs_i=1\}$, $\cB=\{i: \rvs_i=2\}$ and $\cC= \{i: \rvs_i=0\}$. We illustrate the technique when  $|\cA|=|\cB|=|\cC|=\frac{n}{3}$, which roughly corresponds to the case when $\eps=1/3$.  We will use the notation $\rvz_{\cC}^n$ to denote the projection of $\rvz^n$ onto the indices $i \in \cC$ etc. 

In our upper bound, we first reveal the knowledge of $\cB$ to the two encoders non-causally. However the encoders are not aware of the sets $\cA$ and $\cC$. Note from~\eqref{eq:chModel} that $\rvz_{\cB}^n = \rvy_{\cB}^n$, $\rvz_{\cA}^n=\rvx_{\cA}^n$ and $\rvz^n_\cC = \rvx^n_\cC \oplus \rvy^n_\cC$.

\subsubsection{Independence of Input Signals from $\rvw_1 \oplus \rvw_2$}
Observe that $\rvy_{\cB}^n$ is  sub-sequence transmitted by user 2 and hence independent of $\rvu = \rvw_1 \oplus \rvw_2$.  Using this property we have:
{\allowdisplaybreaks \begin{align}
 nR &= H(\rvu)\\
 &= H(\rvu| \rvy_\cB^n)\\
 &=H(\rvu | \rvy_\cB^n, \rvx_\cA^n, \rvz_\cC^n)  + I(\rvx_\cA^n, \rvz_\cC^n; \rvu | \rvy_{\cB}^n)\\
 &\le n(1-\eps) - H(\rvx_\cA^n, \rvz_{\cC}^n|\rvy_{\cB}^n, \rvu)+n\cdot o_n(1), \label{eq:bnd1}
 \end{align}}
where we use Fano's inequality in ${\frac{1}{n}H(\rvu|\rvx_\cA^n, \rvy_\cB^n, \rvz_\cC^n)\le o_n(1)}$ and $o_n(1)$ denotes  a vanishing function in $n$.

\subsubsection{Compound MAC Channel}
 Observe that the same coding scheme must also work when the positions of sets $\cA$ and $\cC$ are interchanged. This results in
 \begin{align}
 nR \le n(1-\eps) - H(\rvx_\cC^n, \rvz_{\cA}^n|\rvy_{\cB}^n, \rvu)+ n\cdot o_n(1). \label{eq:bnd2}
 \end{align}

 Combining~\eqref{eq:bnd1} and~\eqref{eq:bnd2} and ignoring the $o_n(1)$ term, we obtain the following:
{\allowdisplaybreaks \begin{align}
&nR\le n(1-\eps) - \max\bigg(H(\rvx_\cA^n, \rvz_{\cC}^n|\rvy_{\cB}^n, \rvu),H(\rvx_\cC^n, \rvz_{\cA}^n|\rvy_{\cB}^n, \rvu)\bigg) \\
&\le n(1-\eps) - \frac{1}{2}\bigg(H(\rvx_\cA^n, \rvz_{\cC}^n|\rvy_{\cB}^n, \rvu)\!+\!H(\rvx_\cC^n, \rvz_{\cA}^n|\rvy_{\cB}^n, \rvu)\bigg)\\
&\le n(1-\eps) - \frac{1}{2}H(\rvx_\cA^n, \rvz_{\cC}^n,\rvx_\cC^n, \rvz_{\cA}^n|\rvy_{\cB}^n, \rvu)\\
&= n(1-\eps) - \frac{1}{2}H(\rvx_\cA^n, \rvy_{\cC}^n,\rvx_\cC^n, \rvy_{\cA}^n|\rvy_{\cB}^n, \rvu)\\
&\le n(1-\eps) - \frac{1}{2}H(\rvy_{\cA}^n,\rvy_{\cC}^n|\rvy_{\cB}^n, \rvu)\\
&\le n(1-\eps) - \frac{1}{2}H(\rvy_{\cA}^n, \rvy_{\cC}^n|\rvy_{\cB}^n)\label{eq:y_bnd} \end{align}}
 where~\eqref{eq:y_bnd} follows from the fact that the transmit sequence by user $2$, $\rvy^n$ is independent of $\rvw_1$ and hence $\rvw_1\oplus\rvw_2$.
 Eq.~\eqref{eq:y_bnd} suggests that for the rate to be high $(\rvy_{\cA}^n,\rvy_{\cC}^n)$ and $\rvy_{\cB}^n$ must be strongly correlated. However as we show below, such a constraint can only reduce the upper bound obtained by revealing one of the messages to the destination.  
 
 \subsubsection{Penalty from Repetition Coding}
Suppose that the sequence $\rvx^n$ is completely revealed to the destination. The receiver only needs to compute $\rvw_2$ and hence we have:
 \begin{align}
 nR &\le H(\rvy^n) = H(\rvy_\cA^n, \rvy_\cC^n|\rvy_\cB^n) + H(\rvy_{\cB}^n)\label{eq:y_bnd2}
 \end{align}
 Eliminating the joint entropy term between~\eqref{eq:y_bnd} and~\eqref{eq:y_bnd2} we get
 \begin{align}
\frac{3}{2} nR &\le \frac{1}{2}H(\rvy_{\cB}^n) + n(1-\eps) 
 \end{align}
 By using the simple upper bound $H(\rvy_\cB^n) \le |\cB| = n\eps$ we get $R \le \frac{2-\eps}{3}$ which agrees with~\eqref{eq:c_ub} for $\eps=1/3$. 

\subsection{Causal State Feedback}
Consider the case when the encoders are revealed the state sequences in a strictly causal manner. 
The encoding functions at time $i$ can depend on the state sequence up to time $i-1$ 
i.e. $\rvx_i = f_i(\rvw_1, \rvs_1^{i-1})$ and $\rvy_i = g_i(\rvw_2, \rvs_1^{i-1})$.
\begin{prop}
The modulo-sum capacity the multiple access channel with strictly causal state feedback is  lower and upper bounded by $R_\mrm{FB}^- \le C \le R_\mrm{FB}^+$, where
\begin{align}
R_\mrm{FB}^{-} &= \frac{1}{1+2\eps}. \label{eq:c_fb}\\
R_\mrm{FB}^+ &=  1-\eps
\end{align}
\label{prop:c_fb}
\end{prop}

\vspace{-1em}

The lower bound is achieved by a two-phase protocol where the users transmit uncoded bits in the first phase and use a multiple-access code in the second phase.
The upper bound is the genie-aided bound where one of the messages is revealed to the destination. The problem reduces to communicating the other message, say $\rvw_2$ 
to the destination.  Feedback in such a case is well known to not increase the point-to-point capacity.

\subsection{ Numerical Comparisons}

\begin{figure*}
\centering
\includegraphics[scale=0.4]{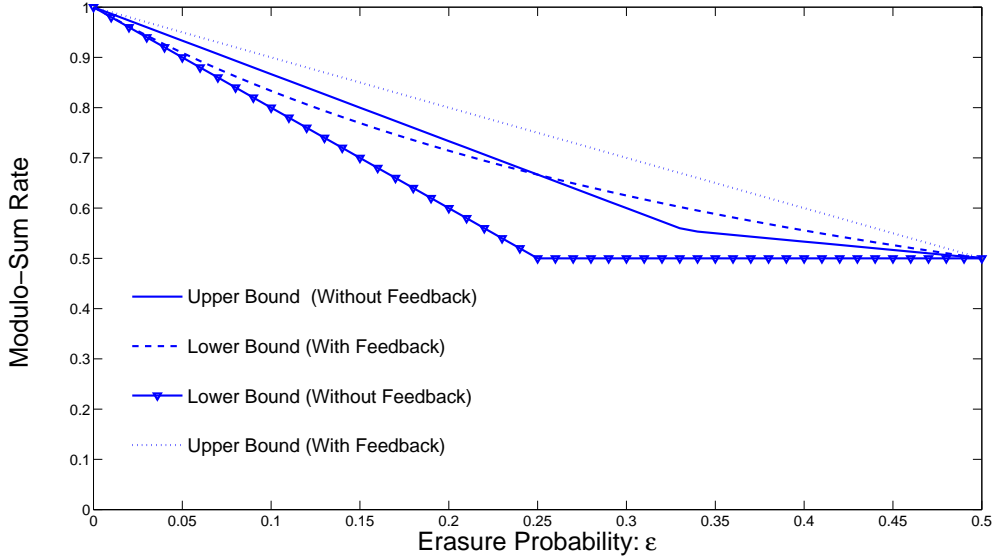}
\caption{Comparison of upper and lower bounds for the Erasure-MAC channel with and without feedback.}
\label{fig:comp}
\end{figure*}

Fig.~\ref{fig:comp} provides a numerical computation of the upper and lower bounds for the Erasure MAC channel both with and without feedback. The upper-most dotted curve corresponds to $R^+_\mrm{FB}=1-\eps$ and is the upper bound on the capacity with feedback. The lowermost curve, marked with backward arrows, is the lower bound achieved by either  the decode and forward or the compute and forward schemes. The other solid curve is our new upper bound on the capacity without feedback (c.f. Theorem~\ref{thm:c_ub}). The fourth curve is the lower bound with feedback in Prop.~\ref{prop:c_fb}. Interestingly we see that it lies above the upper bound for certain values of $\eps$, thus establishing that feedback helps in computation over the erasure multiple access channel.

\subsection{Lossy Reconstruction}

While the focus of this paper is on lossless recovery, our  ideas can be also extended to lossy recovery. We illustrate this with one example.
As before we consider the case when the two transmitters observe i.i.d.\ equiprobable binary sequences $\rvb_1^k$
and $\rvb_2^k$ respectively. The receiver is interested in the modulo-sum  $\rvu^k = \rvb_1^k \oplus \rvb_2^k$. However it suffices to output any sequence
$\hat{\rvu}^k$ that satisfies the distortion constraint
\begin{align}
E\left[ \frac{1}{k}\sum_{i=1}^k \rho(\rvu_i, \hat{\rvu}_i) \right] \le D \label{eq:D-def}
\end{align}
where $\rho(\cdot, \cdot)$ is the associated distortion measure. In this paper we select the erasure distortion measure i.e.,
\begin{align}
\rho(\rvu, \hat{\rvu}) = \begin{cases}
0, &  \hat{\rvu} = \rvu\\
1, & \hat{\rvu} = \star\\
\infty, & \text{ otherwise}
\end{cases}
\label{eq:dist-funcn}
\end{align}
We assume a bandwidth expansion factor of $\beta$. Thus the number of channel uses is $n = k\beta$ and the transmitters generate $\rvx_i^n = f_k(\rvb_i^k)$ for $i=1,2$
and the receiver outputs $\hat{\rvu}^k = g_k(\rvz^n, \rvs^n)$. A distortion $D$ is achievable  if there exist a sequence of encoding 
and decoding functions that satisfy~\eqref{eq:D-def} as $k \rightarrow \infty$.  We develop bounds on the achievable distortion.
\begin{thm}
\label{thm:lossy}
An achievable distortion for modulo-sum reconstruction  of equiprobable and independent binary sources over the erasure multiple access channel satisfies $D_\mrm{outer} \le D \le D_\mrm{inner}$ where
\begin{align}
D_\mrm{inner} &= (1-\beta R^-)^+ \label{eq:D_inner}\\
D_\mrm{outer} &= \left(1- \beta R^+\right)^+ \label{eq:D_outer}
\end{align}
where $R^-$ and $R^+$ are the lower and upper bounds on the modulo-sum capacity stated in  \eqref{eq:c_lb} and \eqref{eq:c_ub} respectively and the function $(v)^+$ equals zero if $v < 0$ and equals $v$ otherwise.
\end{thm}

In particular, examining the expression for $D_\mrm{inner}$ it can be shown that uncoded transmission is sub-optimal even when $\beta =1$ i.e., there is no bandwidth mis-match.
If the two users select $\rvx_i^n = \rvs_i^n$ for $i=1,2$ then the destination must declare an erasure whenever $\rvs_i \neq 0$. It is easy to see that the average distortion 
for this technique equals $2\eps$. In contrast the expression~\eqref{eq:D_inner} equals $\min(2\eps, \frac{1}{2})$ when $\beta =1$. This is a strict improvement for $\eps \in \left(\frac{1}{4}, \frac{1}{2}\right)$.

\subsection{Extended Multiple Access Channel}

We consider an extension of the model in~\eqref{eq:chModel} where when there are two additional states --- either the decoder observes both $(\rvx, \rvy)$ or it observes an erasure. In particular we have that, $\rvs \in \{0,1,2,3,4\},$ where 
\begin{align}
\rvz =\begin{cases}
\rvx \oplus \rvy, & \rvs = 0,\\
\rvx, & \rvs = 1,\\
\rvy, & \rvs = 2,\\
(\rvx,\rvy), & \rvs = 3,\\
\star, & \rvs = 4.
\end{cases}\label{eq:chModel-ext}
\end{align}

Our upper and lower  bounds can be naturally extended to the extended multiple access channel~\eqref{eq:chModel-ext}.
For simplicity we only focus on the lossless case.
Let $\Pr(\rvs=1) = \Pr(\rvs=2)= \del \cdot \eps$, $\Pr(\rvs=0) = \del(1-2\eps)$,  $\Pr(\rvs=3) = \g$ and $\Pr(\rvs=4) =1-\g-\del$.

\begin{prop}
 The modulo-sum capacity of the extended multiple access channel in~\eqref{eq:chModel-ext} satisfies $R^- \le C \le R^+$, where:
\begin{align}
R^- &= \g + \del\cdot\max\left(\frac{1}{2}, (1-2\eps)\right)\label{eq:RLB:Full}\\
R^+ &= \g + \del\left(\frac{2-\eps  + (1-3\eps)^+}{3}\right)\label{eq:RUB:Full}
\end{align}
\label{prop:E-MAC}
\end{prop}

We observe that the lower and upper bounds for the extended model reduce to the corresponding bounds for the simplified model when $\g=0$
and $\del=1$.

\section{ Lower Bound: Proof of Prop.~\ref{prop:c_lb}}

We separately establish the achievability of $R = 1-2\eps$ and $R =1/2$.
\subsection{Compute and Forward Scheme}
We use identical linear codebooks at the two transmitters in the compute and forward scheme to achieve ${R = 1-2\eps}$.
Recall that the messages $\bw_1$ and $\bw_2$ are assumed to be binary valued sequences of length $nR$ bits i.e., we take 
\begin{equation}
\bfb_i^T = \left[b_{i1}, \ldots, b_{i K}\right]
\end{equation}
where $K = nR$ denote the number of information bits in the message.  Let $G$ be a
matrix of dimensions $K \times n$, and let each entry in $G$ be sampled independently from an equiprobable Bernoulli distribution. It is useful to express 
\begin{equation}
\bG = \left[\bg_1, \ldots, \bg_n\right]
\end{equation}
where each $\bg_i \in \{0,1\}^{K}$ is a length $K$ binary valued column vector.  The transmitted sequence $\bx^T = [x_{1},\ldots, x_{K}]$ at receiver $1$ is expressed as:\begin{align}
\bx^T &= \bfb_1^T \cdot G \\
&= [\bfb_1^T \bg_1, \ldots, \bfb_1^T \bg_n]
\end{align}
The transmitted sequence $\by^T$ at user $2$ is defined in a similar manner.  

The receiver is interested in computing 
\begin{align}
\bu^T = \bfb_1^T \oplus \bfb_2^T 
= \left[b_{11} \oplus b_{21},\ldots, b_{1K} \oplus b_{2K}\right].
\end{align}
Given  our specific encoder,  the received symbol  can be expressed as:\begin{align}
 \rvz_i = \begin{cases}
 (\bfb_1^T \oplus \bfb_2^T) \bg_i, & \rvs_i = 0,\\
 \bfb_1^T \bg_i, & \rvs_i = 1,\\
 \bfb_2^T \bg_i, & \rvs_i = 2.
 \end{cases}\label{eq:channel_state_model_2}
 \end{align}
Our proposed decoder only uses the output of the channel when $\rvs_i=0$ and declares  erasures if $\rvs_i \neq 0$.
Let $\hat{G}_0 = G_{| \rvs_i = 0}$ be collection of column vectors in $G$ when $\rvs_i=0$. 
We use the following lemma regarding $\hat{G}_0$:
\begin{lemma}
 For every $\del >0$, there exists a function $o_{n, \del}(1)$ that goes to zero as $n\rightarrow \infty$,  such that following holds:
\begin{align}
\Pr\left(\mathrm{rank}(\hat{G}_0) \ge \min(K, n(1-2\eps-\del))\right) \ge 1-o_{n,\del}(1).
\label{eq:rank-G0}
\end{align}\label{lem:rank-G0}
\end{lemma}

The proof of Lemma~\ref{lem:rank-G0} is obtained by showing that, with high probability, each randomly selected column of $\hat{G}_0$ is in a {\em general position}. 
We omit the proof.  Clearly the receiver can uniquely recover $(\bfb_1^T \oplus \bfb_2^T)$ from
\begin{align}
\bz_0^T = (\bfb_1^T \oplus \bfb_2^T)\cdot \hat{G}_0
\end{align}
if $\hat{G}_0$ has full row-rank, which holds if $R \le 1-2\eps-\del$.
Since $\del>0$ is arbitrary this establishes our first lower bound.

\subsection{Achievability of $R=1/2$: Decode and Forward Approach}
The rate $R=1/2$ is achieved by transmitting both $\rvw_1$ and $\rvw_2$ to the destination instead of taking advantage of the fact that the destination only requires $\rvw_1 \oplus \rvw_2$. 
The multiple access capacity region is given by the convex hull of rate pairs $(R_1, R_2)$ that satisfy:
\begin{align}
R_1 &\le I(\rvx; \rvz, \rvs|\rvy) \label{eq:Mac-1}\\
R_2 &\le I(\rvy; \rvz,\rvs|\rvx)\label{eq:Mac-2}\\
R_1 + R_2 &\le I(\rvx, \rvy; \rvz,\rvs)\label{eq:Mac-3}
\end{align}

Taking $\rvx$ and $\rvy$ to be independent equiprobable binary symbols we get that MAC Capacity region contains $R_1 \le 1-\eps$, $R_2\le 1-\eps$ and $R_1+R_2\le 1$. Since $\eps < 1/2$ the rate pair $R_1 = R_2 = \frac{1}{2}$ is achievable. Thus each user can transmit $\rvw_i$ at a rate of $R=1/2$ to the destination. The destination then computes $\rvw_1 \oplus \rvw_2$.

\begin{remark}
The rate $R=1/2$ can be achieved using a decode and forward scheme even when the two transmitters use  identical codebooks. As established in~\cite{hern}, in addition to~\eqref{eq:Mac-1}-\eqref{eq:Mac-3},
an additional constraint $$R \le I(\rvx, \rvy; \rvz, \rvs| \rvx \oplus \rvy) = 2\eps$$ must be satisfied when identical codebooks are used.   Thus the achievable rate now reduces to $R = \min(1/2, 2\eps)$. Note that with with identical codebooks, the rate $R=1/2$ is achievable for $\eps >1/4$, the region in which decode and forward dominates compute and forward discussed before.
\end{remark}

\subsection{Achieving $R=1/2$ with Compute and Forward}

The rate $R=1/2$ can also be achieved using identical linear codes if the receiver does not ignore the output when $\rvs_i \neq 0$. Let 
Let $\hat{G}_0 = G_{|\rvs_i=0}$,  $\hat{G}_1= G_{|\rvs_i=1}$ and $\hat{G}_2 = G_{|\rvs_i=2}$ be the projections of $G$ onto the indices where $\rvs_{i}=0$, $\rvs_i=1$ and $\rvs_i=2$ respectively.  Following~\eqref{eq:channel_state_model_2}, we let $\bz_\cC^T = (\bfb_1^T +\bfb_2^T)\hat{G}_0$, $\bz_\cA^T =\bfb_1^T \hat{G}_1$ and $\bz_\cB^T = \bfb_2^T \hat{G}_2$.  Furthermore along the lines of Lemma~\ref{lem:rank-G0}, it follows that for any $\del >0$, with a probability that exceeds $1-o_{n,\del}(1)$, we have that
\begin{align}
\mrm{dim}\left(\mrm{col\text{-}space}({\hat{G}_1}) \cup \mrm{col\text{-}space}({\hat{G}_2}) \right) &\le n\cdot \min(2\eps +\del, R)
\end{align}
and since the columns of ${\hat{G}_i}$ are independently sampled, it follows that,
\begin{align}
\mrm{dim}\left(\mrm{col\text{-}space}({\hat{G}_i})\right) &\ge n\cdot \min(\eps - \frac{\del}{2}, R), \quad i=1,2.
\end{align}
Thus using the relation
\begin{align}
\!\!&\mrm{dim}\left(\mrm{col\text{-}space}({\hat{G}_1}) \cap \mrm{col\text{-}space}({\hat{G}_2}) \right)\!\! =
\mrm{dim}\left(\mrm{col\text{-}space}({\hat{G}_1})\right) \notag\\&\!\!+\!\! \mrm{dim}\left(\mrm{col\text{-}space}({\hat{G}_2})\right)
\!\!-\!\! \mrm{dim}\left(\mrm{col\text{-}space}({\hat{G}_1}) \cup \mrm{col\text{-}space}({\hat{G}_2}) \right)
\end{align}it follows that with a probability that exceeds $1-o_{n,\del}(1)$, we have that
\begin{align}
\mrm{dim}\left(\mrm{col\text{-}space}({\hat{G}_1}) \cap \mrm{col\text{-}space}({\hat{G}_2}) \right) &\ge n\cdot d_{12} \notag\\ &\stackrel{\Delta}{=} n \left(2\eps-R-\del\right)^+\label{eq:dim-bnd}
\end{align}
Thus one can find a matrices $M_{i}$  such that 
\begin{equation}
\hat{G}_1 M_1 = \hat{G}_2 M_2 = A \label{eq:align}
\end{equation}
where $A$ is a full-matrix of dimension $n \times d_{12}$. The receiver first computes
\begin{align}
(\bz_\cA^T \oplus \bz_\cB^T)M &= (\bfb_1^T \oplus \bfb_2^T)\cdot A
\end{align}
and then needs to compute $\bfb_1 \oplus \bfb_2$ from $(\bfb_1 \oplus \bfb_2)^T[\hat{G}_0~~A]$. Since the entries in $\hat{G}_0$ and $A$ are independent the rank of $[\hat{G}_0~~A]$ is, with high probability at-least $n(d_{12} + 1-2\eps -\del)$. From~\eqref{eq:dim-bnd} we can show that $R = \max(\frac{1}{2},1-2\eps)$ is achievable.

\section{ Upper Bound: Proof of Theorem~\ref{thm:c_ub}}
\label{sec:UB}
We begin with some notation. For a given sequence $s^n$ $\cA(s^n) = \{i: s_i = 1\}$ and $\cB(s^n)=\{i: s_i = 2\}$. Let $\cC(s^n) = \{i: s_i =0\}$. 
Define $\rvx^n_{\cA(s^n)}$ to be the projection of the sequence $\rvx^n$ on the indices where $s_i = 1$ and use a similar notation for other indices. 

Since the receiver decodes $\rvu = \rvw_1 \oplus \rvw_2$ from its output, from Fano's inequality, we have that
\begin{align}
\frac{1}{n}H\left(\rvu~|~\rvs^n, \rvz^n\right) \le \del_n \label{eq:fano}
\end{align}
for some sequence $\delta_n$ that goes to zero as $n\rightarrow\infty$.

Now consider
\begin{align}
nR&= H(\rvu)\\
&= H(\rvu|\rvs^n) \label{eq:s_indep}\\
&= H(\rvu|\rvs^n, \rvy^n_{\cB(\rvs^n)})\label{eq:y_indep}\\
&= n\del_n + I(\rvu; \rvx_{\cA(\rvs^n)}^n,\rvz^n_{\cC(\rvs^n)} |\rvs^n, \rvy^n_{\cB(\rvs^n)} )\label{eq:fano_app}\\
&= n\del_n + H(\rvx_{\cA(\rvs^n)}^n,\rvz^n_{\cC(\rvs^n)} |\rvs^n, \rvy^n_{\cB(\rvs^n)}) \notag\\ &\qquad-H(\rvx_{\cA(\rvs^n)}^n,\rvz^n_{\cC(\rvs^n)} |\rvs^n, \rvy^n_{\cB(\rvs^n)},\rvu) \label{eq:ub_twoterms}
\end{align}
where~\eqref{eq:s_indep} follows from the fact that the message $\rvu$ is independent of the sequence $\rvs^n$.
Eq.~\eqref{eq:y_indep} follows from the fact that $\rvu = \rvw_1 \oplus \rvw_2$ is independent of $\rvw_2$ and hence also independent of $\rvy^n$.
Eq.~\eqref{eq:fano_app} follows from the chain rule of mutual information and the application of Fano's inequality.

We upper bound the first entropy term in~\eqref{eq:ub_twoterms} as follows.
\begin{align}
& H(\rvx_{\cA(\rvs^n)}^n,\rvz^n_{\cC(\rvs^n)} |\rvs^n, \rvy^n_{\cB(\rvs^n)}) \le H(\rvx_{\cA(\rvs^n)}^n,\rvz^n_{\cC(\rvs^n)} |\rvs^n), \label{eq:reduce_yb} \\
&\le \sum_{s^n \in \cS^n} \Pr(\rvs^n = s^n) \left(|\cA(s^n)| + |\cC(s^n)|\right)\label{eq:binary_terms}\\
&= n(1-\eps) + n\del_n \label{eq:ub_twoterms_t1}
\end{align}
where~\eqref{eq:reduce_yb} follows from the fact that conditioning reduces entropy. Eq.~\eqref{eq:binary_terms} follows from the fact that both $\rvx^n$ and $\rvz^n$ are binary sequences. Eq.~\eqref{eq:ub_twoterms_t1} follows from the fact that $\rvs^n$ is sampled i.i.d.\ from a distribution with $\Pr(\rvs=0)=1-2\eps$ and $\Pr(\rvs=1) =\Pr(\rvs=2)=\eps$.  

Substituting~\eqref{eq:ub_twoterms_t1} into~\eqref{eq:ub_twoterms} we have:
\begin{align}
nR&\le n(1-\eps) + n\del_n - H(\rvx_{\cA(\rvs^n)}^n,\rvz^n_{\cC(\rvs^n)} |\rvs^n, \rvy^n_{\cB(\rvs^n)},\rvu)\label{eq:simplified_entropy_term}
\end{align}

We now separately consider the cases when either $0\le\eps < \frac{1}{3}$ and when $\frac{1}{3} < \eps \le \frac{1}{2}$
\subsection{Case: $\frac{1}{3} < \eps \le \frac{1}{2}$}
Let $\cT_n \subset \cS^n$ be the set of all sequences such that $$|\cA(s^n)| > |\cC(s^n)|.$$
By the weak law of large numbers we have that $\Pr(\rvs^n \in \cT_n) \ge 1-\del_n$ and $\Pr(\rvs^n \in \cT_n) \le \del_n$ for some sequence $\del_n$ that approaches zero as $n\rightarrow \infty$.

For each $s^n \in \cT_n$ we define a permutation function as follows. Let $\cA_1(s^n)$ denotes the first $|\cC(s^n)|$ indices of $s^n$ where $s_i=1$ and $\cA_2(s^n)$ denotes the remaining indices. 
Thus $\cA(s^n) = \cA_1(s^n) \cup \cA_2(s^n)$ and every element in $\cA_1(s^n)$ is smaller than every element of $\cA_2(s^n)$. The permutation function $\pi(s^n)$ is chosen such that $\cC(\pi(s^n)) = \cA_1(s^n)$ and $\cA_1(\pi(s^n)) = \cC(s^n)$. Furthermore $\cA_2(\pi(s^n)) = \cA_2(s^n)$ and $\cB(\pi(s^n)) = \cB(s^n)$. Note that $|\cA(s^n)| = |\cA(\pi(s^n))|$, $|\cB(s^n)| = |\cB(\pi(s^n))|$ and $|\cC(s^n)| = |\cC(\pi(s^n))|$ holds.  Furthermore since the probability of each sequence only depends on its type, we have $\Pr(\rvs^n = s^n) = \Pr(\rvs^n = \pi(s^n))$ for each $s^n \in\cT_n$.

Observe that for each $\rvs^n = s^n \in \cT_n$ we have that,{\allowdisplaybreaks{\begin{align}
&\!\!H(\rvx_{\cA(s^n)}^n,\rvz^n_{\cC(s^n)} |\rvy^n_{\cB(s^n)},\!\rvu,\rvs^n)\notag\\ &\qquad+\!\!H(\rvx_{\cA(\pi(s^n))}^n,\rvz^n_{\cC(\pi(s^n))} |\rvy^n_{\cB(\pi(s^n))},\!\rvu,\rvs^n)\\
&=H(\rvx_{\cA_1(s^n)}^n,\rvx_{\cA_2(s^n)}^n,\rvz^n_{\cC(s^n)} |\rvy^n_{\cB(s^n)},\!\rvu,\rvs^n)\!\!\notag\\
&\qquad + H(\rvx_{\cA_1(\pi(s^n))}^n,\rvx_{\cA_2(\pi(s^n))}^n,\rvz^n_{\cC(\pi(s^n))} |\rvy^n_{\cB(\pi(s^n))},\!\rvu,\rvs^n)\\
&=H(\rvx_{\cA_1(s^n)}^n,\rvx_{\cA_2(s^n)}^n,\rvz^n_{\cC(s^n)} |\rvy^n_{\cB(s^n)},\!\rvu,\rvs^n)\!\!\notag\\
&\qquad + H(\rvx_{\cC(s^n)}^n,\rvx_{\cA_2(s^n)}^n,\rvz^n_{\cA_1(s^n)} |\rvy^n_{\cB(s^n)},\!\rvu,\rvs^n)\label{eq:constr-pi-case2}\\
&\ge H(\rvx_{\cA_1(s^n)}^n,\rvx_{\cA_2(s^n)}^n,\rvz^n_{\cC(s^n)}, \rvx_{\cC(s^n)}^n,\rvz^n_{\cA_1(s^n)}|\rvy^n_{\cB(s^n)}\!\rvu,\rvs^n)\label{eq:cond-pi-case2}\\
&\ge H(\rvy_{\cA_1(s^n)}^n, \rvy_{\cC(s^n)}^n| \rvy_{\cB(s^n)}^n,\rvs^n)\label{eq:uIndep-pi-case2}
\end{align}}}
where~\eqref{eq:constr-pi-case2} follows from the construction of the permutation function $\pi(\cdot)$. Eq.~\eqref{eq:cond-pi-case2} follows from the chain rule of the entropy function and the fact that conditioning reduces entropy. Eq.~\eqref{eq:uIndep-pi-case2} follows from the fact that $\rvz^n = \rvx^n \oplus \rvy^n$ and the fact that $(\rvs^n,\rvy^n)$ is independent of $(\rvw_1 \oplus \rvw_2)$.

Now using~\eqref{eq:simplified_entropy_term} and the fact that $\cT_n \subset \cS^n$ we have 
\begin{align}
nR&\le n(1-\eps) + n\del_n \notag \\ &\qquad- \sum_{s^n \in \cT_n}H(\rvx_{\cA(s^n)}^n,\rvz^n_{\cC(s^n)} |\rvy^n_{\cB(s^n)},\rvu,\rvs^n=s^n)\Pr(\rvs^n=s^n). \label{eq:simplified_entropy_term-case2}
\end{align}
Similarly applying~\eqref{eq:simplified_entropy_term} to the permuted sequence $\pi(\rvs^n)$ we have
\begin{align}
&nR\le n(1-\eps) + n\del_n -\notag \\ & \sum_{s^n \in \cT_n}H(\rvx_{\cA(\pi(s^n))}^n,\rvz^n_{\cC(\pi(s^n))} |\rvy^n_{\cB(\pi(s^n))},\rvu,\rvs^n=s^n)\Pr(\rvs^n=s^n). \label{eq:simplified_entropy_term-case3}
\end{align}
Combining~\eqref{eq:simplified_entropy_term-case2} and~\eqref{eq:simplified_entropy_term-case3} we have that
\begin{align}
&nR\le n(1-\eps) + n\del_n \notag \\ &- \frac{1}{2}\sum_{s^n \in \cT_n}\bigg\{H(\rvx_{\cA(\pi(s^n))}^n,\rvz^n_{\cC(\pi(s^n))} |\rvy^n_{\cB(\pi(s^n))},\rvu,\rvs^n=s^n)\notag\\ & + H(\rvx_{\cA(s^n)}^n,\rvz^n_{\cC(s^n)} |\rvy^n_{\cB(s^n)},\rvu,\rvs^n=s^n)\bigg\}\Pr(\rvs^n=s^n)\\
&\le n(1-\eps) + n\del_n -\notag\\& \frac{1}{2}\sum_{s^n \in\cT_n} H(\rvy_{\cA_1(s^n)}^n, \rvy_{\cC(s^n)}^n| \rvy_{\cB(s^n)}^n,\rvs^n =s^n)\Pr(\rvs^n=s^n)\label{eq:tosub}
\end{align}
where the last relation follows from~\eqref{eq:uIndep-pi-case2}. Now observe that:
\begin{align}
& \sum_{s^n \in \cT_n^c} H(\rvy_{\cA_1(s^n)}^n, \rvy_{\cC(s^n)}^n| \rvy_{\cB(s^n)}^n,\rvs^n =s^n)\Pr(\rvs^n=s^n)\\
&\le \sum_{s^n \in \cT_n^c} n\Pr(\rvs^n=s^n) \le n \del_n \label{label:atyp_bound}
\end{align}where the second step follows from the fact that the sequence $\rvy^n$ is binary valued and the last step follows from the fact that $\Pr(\rvs^n \in \cT_n) \ge 1-\del_n$ holds.
Now observe that
\begin{align}
&\sum_{s^n \in\cT_n} H(\rvy_{\cA_1(s^n)}^n, \rvy_{\cC(s^n)}^n| \rvy_{\cB(s^n)}^n,\rvs^n =s^n)\Pr(\rvs^n=s^n)\\
&= \sum_{s^n \in\cS_n} H(\rvy_{\cA_1(s^n)}^n, \rvy_{\cC(s^n)}^n| \rvy_{\cB(s^n)}^n,\rvs^n =s^n)\Pr(\rvs^n=s^n) \notag\\
&- \sum_{s^n \in \cT_n^c} H(\rvy_{\cA_1(s^n)}^n, \rvy_{\cC(s^n)}^n| \rvy_{\cB(s^n)}^n,\rvs^n =s^n)\Pr(\rvs^n=s^n) \\
&\ge H(\rvy^n_{\cA_1(\rvs^n)}, \rvy^n_{\cC(\rvs^n)}|\rvy^n_{\cB(\rvs^n)}, \rvs^n) - n \del_n. 
\end{align}

Substituting into~\eqref{eq:tosub} we arrive at:
\begin{align}
nR&\le n(1-\eps) + 2n\del_n  - \frac{1}{2}H(\rvy^n_{\cA_1(\rvs^n)}, \rvy^n_{\cC(\rvs^n)}|\rvy^n_{\cB(\rvs^n)}, \rvs^n)\\
&\le n(1-\eps) + 2n\del_n  - \frac{1}{2}H( \rvy^n_{\cC(\rvs^n)}|\rvy^n_{\cB(\rvs^n)}, \rvs^n). \label{eq:lower_bnd_t1}
\end{align} Also since the decoder is able to compute $\rvw_1 \oplus \rvw_2$ from  $(\rvz^n,\rvs^n)$, we have:
{\allowdisplaybreaks \begin{align}
nR&= H(\rvw_2 \oplus \rvw_1) \\
&= H(\rvw_2|\rvw_1) \label{eq:indep_msg_case2}\\
&=H(\rvw_2|\rvw_1,\rvs^n) \label{eq:indep_state_case2}\\
&=H\left(\rvw_2|\rvw_1,\rvs^n,\rvx^n_{\cA(\rvs^n)},\rvx^n_{\cC(\rvs^n)}\right)\label{eq:indep_za_case2}\\
&=H\left(\rvw_2|\rvw_1,\rvs^n,\rvx^n_{\cA(\rvs^n)},\rvx^n_{\cC(\rvs^n)},\rvy_{\cB(\rvs^n)}^n, \rvy_{\cC(\rvs^n)}^n\right)\notag\\ &\qquad + I(\rvw_2; \rvy_{\cB(\rvs^n)}^n, \rvy_{\cC(\rvs^n)}^n|\rvw_1,\rvs^n,\rvx^n_{\cA(\rvs^n)},\rvx^n_{\cC(\rvs^n)})\\
&\le n\del_n + H(\rvy_{\cB(\rvs^n)}^n, \rvy_{\cC(\rvs^n)}^n|\rvs^n) \label{eq:fano_app_case2_2}\\
&= n\del_n + H(\rvy_{\cC(\rvs^n)}^n |\rvs^n,\rvy_{\cB(\rvs^n)}^n) + H(\rvy_{\cB(\rvs^n)}^n|\rvs^n)\label{eq:cutset_bound_case_2}
\end{align}}
where~\eqref{eq:indep_msg_case2} follows from the fact that $\rvw_1$ and $\rvw_2$ are independent. Eq.~\eqref{eq:indep_state_case2} follows from the fact
that the state sequence is independent of $(\rvw_1, \rvw_2)$.Eq.~\eqref{eq:indep_za_case2} follows from the fact that 
from construction, $(\rvx_{\cA(\rvs^n)}^n,\rvx_{\cC(\rvs^n)}^n)$ consists entirely of symbols transmitted by user $1$ and hence is independent of $\rvw_2$. Finally, Eq.~\eqref{eq:fano_app_case2_2} follows by applying Fano's inequality since $\rvw_1 \oplus \rvw_2$ can be decoded from $(\rvz^n,\rvs^n)$.
Combining~\eqref{eq:lower_bnd_t1} and~\eqref{eq:cutset_bound_case_2} we have that
\begin{align}
\frac{3}{2}R &\le (1-\eps) + \frac{5}{2}\delta_n + \frac{1}{2} E\left[\frac{1}{2n}|\cB(\rvs^n)|\right]\\
&= 1-\frac{1}{2}\eps + \frac{5}{2}\del_n.
\end{align}
Since $\del_n$ vanishes to zero as $n\rightarrow\infty$ we recover $R \le \frac{2-\eps}{3}$ as required. 

\subsection{Case: $0 \le \eps < \frac{1}{3}$}
We let $\cT_n\subseteq \cS^n$ to be the set of all sequences such that $|\cC(s^n)| > |\cA(s^n)|$. From the weak law of large numbers we have that
$\Pr(\rvs^n \in \cT_n) \ge 1- \del_n$ and $\Pr(\rvs^n \notin \cT_n) \le \del_n$, for some sequence $\del_n$ that goes to zero as $n\rightarrow\infty$.

Split  the set $\cC(s^n)$ as a union of two sets i.e.,  $\cC(s^n) = \cC_1(s^n) \cup \cC_2(s^n)$. Let  $\cC_1(s^n)$ be the first $|\cA(s^n)|$ elements of $\cC(s^n)$ 
i.e., $|\cC_1(s^n)| = |\cA(s^n)|$ and each index in $\cC_1(s^n)$ be smaller than each index in $\cC_2(s^n)$. We let $\pi(s^n)$ be a permutation function such that $\cC_1(s^n) = \cA(\pi(s^n))$
and $\cA(s^n) = \cC_1(\pi(s^n))$. Let $\cC_2(s^n) = \cC_2(\pi(s^n))$ and $\cB(s^n) = \cB(\pi(s^n))$.

Following the the sequence of steps similar to~\eqref{eq:uIndep-pi-case2} we have that for each $\rvs^n \in \cT_n$,
\begin{align}
&\!\!H(\rvx_{\cA(s^n)}^n,\rvz^n_{\cC(s^n)} |\rvy^n_{\cB(s^n)},\!\rvu, \rvs^n)\notag\\ &+\!\!H(\rvx_{\cA(\pi(s^n))}^n,\rvz^n_{\cC(\pi(s^n))} |\rvy^n_{\cB(\pi(s^n))},\!\rvu, \rvs^n)\\
&=H(\rvx_{\cA(s^n)}^n,\rvz^n_{\cC_1(s^n)},\rvz^n_{\cC_2(s^n)} |\rvy^n_{\cB(s^n)},\!\rvu, \rvs^n)\notag\\ &\qquad+\!\!H(\rvx_{\cC_1(s^n)}^n,\rvz^n_{\cA(s^n)}, \rvz^n_{\cC_2(s^n)} |\rvy^n_{\cB(s^n)},\!\rvu, \rvs^n)\label{eq:constr_case3}\\
&\ge H(\rvx_{\cA(s^n)}^n,\rvz^n_{\cC_1(s^n)},\rvz^n_{\cC_2(s^n)},\rvx_{\cC_1(s^n)}^n,\rvz^n_{\cA(s^n)}|\rvy^n_{\cB(s^n)},\!\rvu, \rvs^n)\label{eq:cond_case3}\\
&= H(\rvx_{\cA(s^n)}^n,\rvx^n_{\cC_1(s^n)},\rvz^n_{\cC_2(s^n)},\rvy^n_{\cA(s^n)},\rvy_{\cC_1(s^n)}^n|\rvy^n_{\cB(s^n)},\!\rvu, \rvs^n) \label{eq:det_fun_case3}\\
&\ge H(\rvy^n_{\cC_1(s^n)}, \rvy^n_{\cA(s^n)}|\rvy_{\cB(s^n)}^n,\rvu, \rvs^n)\\
&= H\left(\rvy^n_{\cC_1(s^n)}, \rvy^n_{\cA(s^n)}|\rvy_{\cB(s^n)}^n, \rvs^n\right)\label{eq:cond_ent_bnd_case3}
\end{align}
where~\eqref{eq:constr_case3} follows from the construction of the permutation function $\pi(\cdot)$ and the fact that $\cC(s^n) = \cC_1(s^n) \cup \cC_2(s^n)$. Eq.~\eqref{eq:cond_case3} follows from the chain rule of entropy and the fact that conditioning reduces entropy. Eq.~\eqref{eq:det_fun_case3} follows from the fact that $\rvz^n = \rvx^n \oplus \rvy^n$.  Eq.~\eqref{eq:cond_ent_bnd_case3} follows from the fact that $\rvu=\rvw_1 \oplus \rvw_2$ is independent of $\rvw_2$  and hence~$\rvy^n$.
Following the sequence of steps similar to~\eqref{eq:lower_bnd_t1} we have that:
\begin{align}
nR &\le n(1-\eps) + 2n\del_n - \frac{1}{2} H(\rvy^n_{\cA(\rvs^n)}, \rvy^n_{\cC_1(\rvs^n)}|\rvy^n_{\cB(\rvs^n)},\rvs^n) \\
&\le n(1-\eps) + 2n\del_n - \frac{1}{2} H( \rvy^n_{\cC_1(\rvs^n)}|\rvy^n_{\cB(\rvs^n)},\rvs^n). \label{eq:R_ub_case3}
\end{align}

Following the sequence of steps leading to~\eqref{eq:cutset_bound_case_2} we have
\begin{align}
nR&\le  n\del_n + H(\rvy_{\cC_1(\rvs^n)}^n |\rvs^n,\rvy_{\cB(\rvs^n)}^n) + H(\rvy_{\cB(\rvs^n)}^n, \rvy_{\cC_2(\rvs^n)}^n|\rvs^n).
\label{eq:cutset_bound_case_3} 
\end{align}Combining~\eqref{eq:cutset_bound_case_3} and~\eqref{eq:R_ub_case3} we have
\begin{align}
\frac{3}{2}nR &\le \frac{5}{2}n\del_n + n(1-\eps) + \frac{1}{2}H(\rvy_{\cB(\rvs^n)}^n, \rvy_{\cC_2(\rvs^n)}^n|\rvs^n)\\
&\le \frac{5}{2}n\del_n + n(1-\eps) + \frac{1}{2}E[|\cB(\rvs^n)| + |\cC_2(\rvs^n)|]\\
&\le \frac{5}{2}n\del_n + n(1-\eps) + \frac{n}{2}(1-2\eps).
\end{align}
Since $\del_n$ vanishes to zero, as $n\rightarrow\infty$, $R \le \frac{3-4\eps}{3}$ holds, which completes the proof. 

Thus we have established Theorem~\ref{thm:c_ub} for $0\le \eps < 1/3$ and $1/3 < \eps \le 1/2$. For $\eps=1/3$ the upper bound follows by observing that the capacity is monotonically decreasing in $\eps$ and the upper and lower limits to the upper bound function
at $\eps=1/3$ both equal $5/9$.

\section{Coding Technique with Feedback}

We provide a sketch of the achievable rate with feedback stated in Prop.~\ref{prop:c_fb}. We use a two phase protocol. In the first phase encoders $1$ and $2$ transmit $b_{1i}$ and $b_{2i}$ respectively for $i=1,2\ldots,n$. For those indices where $\rvs_i=0$ the receiver obtains $b_{1i} \oplus b_{2i}$. Among the remaining indices users $1$ and $2$ construct $\hat{w}_1 = \{b_{1j}\}_{j:\rvs_j=2}$ and $\hat{w}_2 = \{b_{2j}\}_{j:\rvs_j = 1}$. In the second phase, the messages $\hat{w}_{1j}$ and $\hat{w}_{2j}$ are transmitted to the destination using a multiple access channel code. By computing the capacity region of the associated multiple access channel (c.f.~\eqref{eq:Mac-1}-\eqref{eq:Mac-3}), it can be verified that the number of channel uses in this phase is $\approx 2n\eps$.  Thus the total rate is $\approx \frac{n}{n+2n\eps} = \frac{1}{1+2\eps}$ as required. 

The upper bound is obtained by revealing one of the messages, say $\rvw_1$, to the destination. Thus only $\rvw_2$ needs to be communicated to the receiver. For such a point-to-point problem, it is well known that feedback does not increase the capacity of $C = 1-\eps$. Thus $R^+=1-\eps$ is an  upper bound even when feedback is available to the transmitters.

\section{Lossy Reconstruction}
\label{sec:lossy}

We establish the bounds stated in Theorem~\ref{thm:lossy}. For the achievability scheme, both the users only encode first $k_1 \le k$ source symbols. The encoding functions
at the two users are selected in order to communicate the modulo-sum $\rvu^{k_1} = \rvb_1^{k_1} \oplus \rvb_2^{k_1}$ in a lossless manner.
Thus user $1$ generates $\rvx^n = f_1(\rvb_1^k)$
and user $2$ generates $\rvy^n = f_2(\rvb_2^k)$ where the encoding functions are selected according to either the compute-and-forward
or decode-and-forward schemes discussed previously. It follows that the decoder can recover $\rvu^{k_1}$ with high probability if
$k_1 \le n R^-$ where $R^- = \max\{\frac{1}{2}, 1-2\eps\}$ is our best achievable rate. The decoder declares an erasure for all indices $j \in [k_1+1, k]$.
 The associated distortion per symbol satisfies
 \begin{align}
D_\mrm{inner} &= \frac{(k-k_1)^+}{k}\\
 &= \left(1 - \beta R^-\right)^+.
 \end{align}
as required. 
For establishing an outer bound on the achievable distortion we note that applying  rate-distortion theorem to the erasure distortion metric
and i.i.d.\ equiprobable binary sources, we have~\cite{coverThomas} that $R(D) = 1-D$. Furthermore from the definition of the rate-distortion function note that
if $D$ is an achievable distortion metric then:
\begin{align}
k R(D) &\le I(\rvu^k; \hat{\rvu}^k)\\
&\le I(\rvu^k; \rvz^n, \rvs^n) \label{eq:data-proc}\\
&=I(\rvu^k; \rvz^n | \rvs^n) \label{eq:s-indep}\\
&=I(\rvu^k; \rvx_{\cA(\rvs^n)}^n,  \rvy_{\cB(\rvs^n)}^n,  \rvz_{\cC(\rvs^n)}^n | \rvs^n) \label{eq:chan-struct}\\
&=I(\rvu^k; \rvx_{\cA(\rvs^n)}^n,   \rvz_{\cC(\rvs^n)}^n | \rvs^n, \rvy_{\cB(\rvs^n)}^n) \label{eq:chan-struct2}\\
&\le nR^+
\end{align}
where~\eqref{eq:data-proc} follows from the data processing theorem and~\eqref{eq:s-indep}
follows from the fact that the source sequences are independent of the state of the channel,
\eqref{eq:chan-struct} follows from the structure of the channel where the sets $\cA(\rvs^n)$,
$\cB(\rvs^n)$ and $\cC(\rvs^n)$ are defined in the beginning of Section~\ref{sec:UB} and~\eqref{eq:chan-struct2}
follows from the fact that $\rvy_{\cB(\rvs^n)}^n$ is a subsequence of the codeword $\rvy^n$ transmitted by user $2$
which is independent of $\rvs_1^k$ and hence $\rvu^k = \rvs_1^k \oplus \rvs_2^k$, since the sequences are i.i.d.\ and
equiprobable. Applying the same steps as in our upper bound (c.f.~\eqref{eq:fano_app}) we have that
\begin{align}
R^+ = \frac{(1-3\eps)^+ + 2-\eps}{3} \label{eq:Rub-exp}
\end{align}
Thus we have that
\begin{align}
D_\mrm{outer} \ge (1 - \beta R^+)^+
\end{align}
where $R^+$ is defined via~\eqref{eq:Rub-exp}.

\section{Extended Multiple Access Channel: Proof of Prop.~\ref{prop:E-MAC}}

In this section we establish the upper and lower bounds stated in Prop.~\ref{prop:E-MAC}.
Recall that for the extended model the channel output $\rvz$ can take one of five possible values:
$\Pr(\rvz=\rvx) = \Pr(\rvz = \rvy) = \del\cdot \eps $, $\Pr(\rvz = \rvx \oplus \rvy) = \del(1-2\eps)$,
$\Pr(\rvz = (\rvx, \rvy))= \g$ and $\Pr(\rvz = \star)=1-\del-\g$. 

\subsection{Proof of Lower Bound~\eqref{eq:RLB:Full}}

We first show that $R^- = \frac{1}{2} \del + \g$ is achievable by communicating two independent messages to the receiver each at rate $R^-$.
Recall that any achievable rate pair $(R_1, R_2)$ of the multiple-access channel can be computed via
\begin{align}
R_1 &\le I(\rvx ; \rvz| \rvy, \rvs), \\
R_2 &\le I(\rvy;\rvz|\rvx, \rvs)\\
R_1 + R_2 &\le I(\rvx, \rvy;\rvz|\rvs)
\end{align}
Evaluating for the equi-probable input distribution we have that
\begin{align}
R_1 &\le \del(1-\eps) + \g\\
R_2 &\le \del(1-\eps) + \g\\
R_1 + R_2 &\le \del + 2\g
\end{align}Since $\eps \le 1/2$ it follows that $R_1 = R_2 = \frac{1}{2}\del + \g$ is an achievable rate-pair. This establishes that ${R^-=\frac{1}{2}\del+\g}$ is achievable.

When identical linear codebooks are used for decode and forward, following~\cite{hern} we require an additional constraint on the rate:
\begin{align*}
R \le I(\rvx, \rvy; \rvz, \rvs|\rvx \oplus \rvy) = \g + 2\del\eps
\end{align*}
and hence the achievable rate reduces to $R = \g + \del \min(2\eps, \frac{1}{2})$. As the decode-and-forward scheme only dominates for $\eps > 1/4$, there is no penalty from the additional rate constraint involved from using identical codebooks. 

To establish that $R^- = \g + \del(1-2\eps)$ is also achievable, we use identical linear codebooks at the two transmitters. In particular transmitter $1$ computes $\bx^T = \bb_1^T G$ and transmitter $2$ computes $\by^T = \bb_2^T G$ where the entries of $G \in {\mathbb F}_2^{nR \times n}$ are sampled i.i.d.\ from an equiprobable Bernoulli distribution. The receiver only keeps the output symbols corresponding to $\rvs = 0$ and $\rvs=4$. When $\rvs=4$ it  computes $\rvz = \rvx \oplus \rvy$ from the received pair $(\rvx, \rvy)$.  Thus the total fraction of non-erasures at the receiver is $\g + \del(1-2\eps)$. It can then be shown, as in Prop.~\ref{prop:c_lb} that $R = \g + \del(1-2\eps)$  is achievable.

\subsection{Proof of Upper Bound~\eqref{eq:RUB:Full}}

Our upper bound analysis closely follows the proof of Theorem~\ref{thm:c_ub}. We only illustrate the main points of difference due to the addition of the two extra state values. Following the steps leading to~\eqref{eq:ub_twoterms}, we can show that
\begin{align}
nR &\le no_n(1) + H(\rvx_{\cA(\rvs^n)}^n,\rvz^n_{\cC(\rvs^n)}, \rvx_{\cD(\rvs^n)}^n |\rvs^n, \rvy^n_{\cB(\rvs^n)}, \rvy_{\cD(\rvs^n)}^n) \notag\\ &\qquad-H(\rvx_{\cA(\rvs^n)}^n,\rvz^n_{\cC(\rvs^n)}, \rvx_{\cD(\rvs^n)}^n |\rvs^n, \rvy^n_{\cB(\rvs^n)},\rvu, \rvy_{\cD(\rvs^n)}^n). \label{eq:R_Fano_t1}
\end{align}
where the sets $\cA,$ $\cB$ and $\cC$ are as defined in Section~\ref{sec:UB} and let $\cD(s^n) = \{i: s_i = 3\}$ and $\cE(s^n) = \{i: s_i = 4\}$. 

Through standard arguments we have
\begin{align}
& H(\rvx_{\cA(\rvs^n)}^n,\rvz^n_{\cC(\rvs^n)}, \rvx_{\cD(\rvs^n)}^n |\rvs^n, \rvy^n_{\cB(\rvs^n)}, \rvy_{\cD(\rvs^n)}^n)\\
&\le E\left[|\cA(\rvs^n)|+ |\cC(\rvs^n)| + |\cD(\rvs^n)|\right] = n \del(1-\eps) + n \g.
\end{align}From~\eqref{eq:R_Fano_t1}, dropping the $o_n(1)$ terms to keep the expressions compact, we have
\begin{align}
nR &\le  n \del(1-\eps) + n \g - \notag\\
&H(\rvx_{\cA(\rvs^n)}^n,\rvz^n_{\cC(\rvs^n)}, \rvx_{\cD(\rvs^n)}^n |\rvs^n, \rvy^n_{\cB(\rvs^n)},\rvu, \rvy_{\cD(\rvs^n)}^n). 
\label{eq:Rfull:t1}
\end{align}

We assume that $0 \le \eps < 1/3$ and let $\cT_n$ denote all sequences $s^n$ such that $|\cC(s^n)|> |\cA(s^n)|$. As before let $\cC(s^n) = \cC_1(s^n )\cup \cC_2(s^n)$ where $\cC_1(s^n)$ denotes the first $|\cA(s^n)|$ elements of $\cC(s^n)$. From the weak law of large numbers $\Pr(\rvs^n \in \cT_n) \ge 1- o_n(1)$ holds. 

Let $\pi(s^n)$ denote a permutation of $s^n$ such that $\cC_1(\pi(s^n)) = \cA(s^n)$ and $\cA(\pi(s^n)) = \cC_1(s^n)$. Furthermore let $\cB(\pi(s^n)) = \cB(s^n)$ and $\cC_2(\pi(s^n)) = \cC_2(s^n)$ be satisfied. Also the sets $\cD$ and $\cE$ are invariant under this permutation mapping.  Applying~\eqref{eq:Rfull:t1} to the sequence $\pi(s^n)$ we have that

\begin{align}
nR&\le n \del(1-\eps) + n \g - \notag\\
&H(\rvx_{\cA(\pi(\rvs^n))}^n,\rvz^n_{\cC(\pi(\rvs^n))}, \rvx_{\cD(\pi(\rvs^n))}^n |\rvs^n, \rvy^n_{\cB(\rvs^n)},\rvu, \rvy_{\cD(\rvs^n)}^n). 
\label{eq:Rfull:t2}
\end{align}

By following the steps leading to~\eqref{eq:cond_ent_bnd_case3} we can show that
\begin{align}
& H(\rvx_{\cA(\rvs^n)}^n,\rvz^n_{\cC(\rvs^n)}, \rvx_{\cD(\rvs^n)}^n |\rvs^n, \rvy^n_{\cB(\rvs^n)}, \rvy_{\cD(\rvs^n)}^n,\rvu)+\notag\\
&H(\rvx_{\cA(\pi(\rvs^n))}^n,\rvz^n_{\cC(\pi(\rvs^n))}, \rvx_{\cD(\pi(\rvs^n))}^n |\rvs^n, \rvy^n_{\cB(\rvs^n)},\rvu, \rvy_{\cD(\rvs^n)}^n)\\
&\ge H(\rvy^n_{\cA(\rvs^n)}, \rvy_{\cC_1(\rvs^n)}^n|\rvs^n, \rvy^n_{\cB(\rvs^n)},\rvu, \rvy_{\cD(\rvs^n)}^n).
\label{eq:Rfull:t3}
\end{align}
It follows from~\eqref{eq:Rfull:t1},~\eqref{eq:Rfull:t2} and~\eqref{eq:Rfull:t3} that
\begin{align}
nR &\le n \del(1-\eps) + n \g - 
H( \rvy_{\cC_1(\rvs^n)}^n|\rvs^n, \rvy^n_{\cB(\rvs^n)}, \rvy_{\cD(\rvs^n)}^n)\label{eq:R_bnd:full}.
\end{align}
Furthermore if $\rvx^n$ is revealed to the decoder, it follows that the decoder must decode $\rvw_2$. Thus
\begin{align}
&nR \le H(\rvy_{\cB(\rvs^n)}^n, \rvy_{\cC(\rvs^n)}^n, \rvy_{\cD(\rvs^n)}^n|\rvs^n)\\
&\!\!= \!\!H(\rvy_{\cB(\rvs^n)}^n, \rvy_{\cD(\rvs^n)}^n,\rvy^n_{\cC_2(\rvs^n)}|\rvs^n) + H(\rvy^n_{\cC_1(\rvs^n)}|\rvs^n,\rvy_{\cB(\rvs^n)}^n, \rvy_{\cD(\rvs^n)}^n)\\
&\le n(\g + \del\eps)+ n(1-3\eps)\del + H(\rvy^n_{\cC_1(\rvs^n)}|\rvs^n,\rvy_{\cB(\rvs^n)}^n, \rvy_{\cD(\rvs^n)}^n)\label{eq:R_bnd:full2}.
\end{align}
Combining~\eqref{eq:R_bnd:full} and~\eqref{eq:R_bnd:full2} to eliminate the entropy term we have that
\begin{align}
\frac{3}{2}nR &\le \frac{3}{2}n\g + n\del(1-\frac{1}{2}\eps)  + \frac{n}{2}(1-3\eps)\del,
\end{align}
which results in 
\begin{align}
R \le \g + \del\left(\frac{2-\eps + (1-3\eps)}{3}\right)
\end{align}
for $\eps < 1/3$. For $\eps > 1/3$, one can  similarly establish that
\begin{align}
R \le \g + \del\left(\frac{2-\eps}{3}\right),
\end{align}which completes the upper bound analysis. 

\section{Conclusions}

We study computation of the modulo-sum of two messages over a multiple access channel with erasures. Unlike the Gaussian  channel model, this model does not have a suitable structure to directly compute the modulo sum. Our main result is an upper bounding technique  that converts the setup to a compound multiple-access channel and results in a tighter upper bound than the usual cut-set bound.
Using this bound we establish that a simple ARQ type feedback can increase the modulo-sum capacity for our channel. We also consider the case when a lossy reproduction of the modulo-sum
is required and observe that uncoded transmission is sub-optimal even when there is no bandwidth mismatch. 

While function-computation over Gaussian networks has recently received a significant attention, the problem is far less understood when we consider other  relevant channel models.
We hope that techniques developed in this paper are useful in other related problems in this emerging area. 

\bibliographystyle{IEEEtran}
\bibliography{sm}

\end{document}